\documentstyle[twoside, epsfig]{article}
\input pz.sty

\begin{document}

\PZhead {4}{34}{2014}{December 1}{December 10}

\PZtitle{Explosion of an LBV star in the galaxy UGC~8246}

\PZauth{E.A.~Barsukova$^1$, V.P.~Goranskij$^2$, A.F.~Valeev$^{1,3}$, and
S.S. Kaisin$^1$}

\PZinsto{$^1$}{Special Astrophysical Observatory, Russian Academy
of Science, Nizhnij Arkhyz, Karachai-Cherkesia, 369167, Russia;
e-mail: bars@sao.ru}

\PZinsto{$^2$}{Sternberg Astronomical Institute, Lomonosov Moscow
University, Universitetski pr., 13, Moscow, 199992, Russia}

\PZinsto{$^3$}{Kazan Federal University, Kremlevskaya 18, Kazan, 420008, Russia}

\SIMBADobj{UGC 8246 OT, PSN J13100734+3410514}

\PZabstract{ We present the results of spectroscopy and CCD
photometry of the intermediate-luminosity red transient
PSN~J13100734+3410514 in the galaxy UGC~8246 performed in February
and June 2014 with the Russian 6-meter telescope and SCORPIO
spectral camera. Our CCD photometry was continued with the Special
Astrophysical Observatory's 1-m telescope till November 2014. The
star was discovered in late December 2013 at visual brightness
17\fmm6, which corresponded to the absolute magnitude $M_V = -12\fmm7$,
and was identified as a supernova impostor. Spectra
taken at the visual brightness level of 19\fmm5 show composite
triple profiles in the H$_\alpha$ and H$_\beta$ emission lines. We
explain the main component of the profiles as radiation from a
photoionized extensive gaseous envelope formed by the stellar wind
of the progenitor before the outburst. The other two components
are treated as radiation from bipolar ejecta. In Balmer line
profiles, there is an evidence for a light echo propagating in the
surrounding medium after the outburst. Our spectra contain
emissions of He I, \hbox{Na I,} \hbox{Mg I,} numerous Fe~II
emissions, the strongest of which have P~Cyg profiles. There are
also [O~II], [O~III], and [S~II] emissions of an H~II region
associated with the transient. The emissions of the region are
superimposed on the star spectrum. The light curves show rapid
decline and color reddening. Our observations confirm that the
UGC~8246 transient was an explosion of a high-mass LBV star. }

\vspace{0.5cm}
\section{INTRODUCTION}

Discoveries of a large number of supernovae using networks of
automatic robotic telescopes and preliminary classification of
their spectra fill the maximum-luminosity gap between classical
novae and supernovae. The optical transients reaching in their
outbursts the absolute visual magnitude from $-$8\mm\ to $-$17\mm\
and evolving their spectra to cooler subtypes during outbursts are
called Intermediate-Luminosity Red Transients (ILRTs). This family
of astrophysical objects is heterogeneous and contains luminous
blue variable (LBV) stars with giant eruptions ($\eta$~Car,
low-luminosity SNe 1961V and 1954J with spectra resembling SNe IIn
in outbursts), calcium transients like SN~2008S in NGC~6496 or NGC
300 OT having super-AGB stars as progenitors, and luminous red
novae like V1006/7 in M31, V838~Mon in the Galaxy, and M85
OT2006--1. Van Dyk et al. (2000) called lower-luminosity SNe IIn
``supernova impostors'', but later this term was extended to other
transients with peak luminosities in the gap (Berger et al. 2009;
Smith et al. 2009).

Some LBVs and Galactic red novae turned out to be binaries or
multiple systems. The most promising scenario for luminous red
novae as merging events in binary systems was suggested by Tylenda
\& Soker (2006). Tylenda et al. (2011) observed a merging event in
the contact binary before the outburst of V1309~Sco as a red nova.
On the other hand, Martini et al. (1999) suggested that a red nova
phenomenon is due to a nuclear event in a single star, in which a
slow shock drives the photosphere outward. Barsukova et al. (2014)
specify that this is the phenomenon connected with the adiabatic
expansion of the star envelope after an episodic energy release in
the star center. The release may be due to merging of stellar
nuclei inside a massive common envelope or to instability of a
single stellar nucleus of a young star. $\eta$~Car, the LBV which
experienced a Great Eruption in the mid-1800s accompanied by an
episodic mass-loss event, is known as a binary (Corcoran \&
Ishibashi 2012). The companion is not seen in the spectrum, but it
is found to be an O4--O6 giant having mass of 40--50~$M_\odot$
based on indirect evidence (Mehner et al., 2010). RXTE X-ray
observations show cyclic variability with the 2024-day period that
may be the signature of the star's motion in an elliptic orbit
with $e = 0.9$ and a semi-major axis $a \approx 15 $~AU (Ishibashi
et al. 1999). However, the role of this companion in the evolution
of $\eta$~Car and in the Great Eruption scenario is still unknown.
Kashi \& Soker (2010) and Kashi (2010) suggest that most outbursts
of LBV-type systems are powered by gravitational energy released
from either a vigorous mass transfer process from the evolved
primary star to a main sequence secondary star or a merger of two
stars. So, major LBV eruptions can be triggered by stellar
companions, and in extreme cases, a short-duration event with a
huge mass transfer rate can lead to a bright transient event on
time scales of weeks to months. Alternative hypotheses explain LBV
eruptions either with enhanced stellar wind episodes when a
massive star luminosity sometimes exceeds the Eddington limit, or
with interior explosions due to instabilities in stellar nuclei or
envelopes that may be pulsational instabilities (Owocki \& Shaviv
2012).

Note that one of LBVs in the galaxy NGC~7259, SN 2009ip,
previously went through two typical impostor outbursts that
reached $M_V = -14\mm$ in maxima and then became a true
core-collapse supernova of --18th absolute magnitude in the peak
of brightness (Mauerhan et al. 2013; Margutti et al. 2014).

\PZfig{8cm}{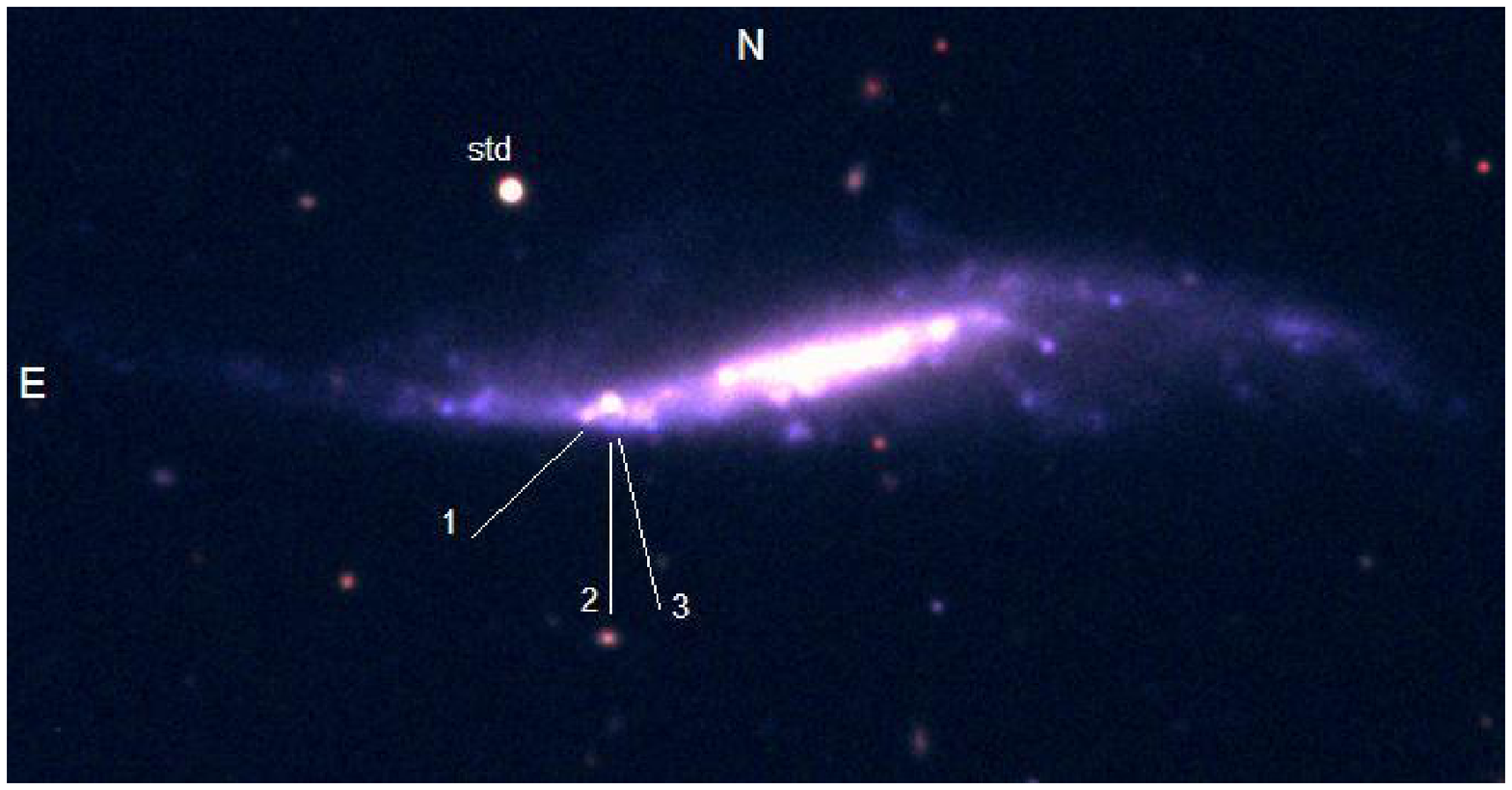}{A color image of the UGC~8246 transient
PSN~J13100734+3410514 composed from BTA/SCORPIO $B$, $V$, and $R$
frames. The image size is $4\arcm\times2\arcm$. The comparison
star is labeled ``std''. The straight lines marked 1, 2, and 3 are
directions of the slit of BTA spectra.}

A recent discovery of an LBV-type transient as a possible
supernova in UGC~8246 on 2013 December 20.93 UT was reported by
Wang and Gao (2013). The object had the $R$~magnitude of
approximately 17.75. UGC~8246 is an SB(s)cd type galaxy, its
redshift is $z = 0.002712 \pm0.000037$ and distance, $11.7
\pm0.8$~Mpc; the Galactic interstellar absorption is $A_V =
0\fmm03$ (NED). With these parameters, the absolute magnitude of
the transient at the time of discovery was $M_V \sim -12\fmm7$.
Additional photometry acquired in 2013 December was published by
Elenin, Wang \& Gao, Luppi \& Buzzi (CBAT) and by Brimacombe
(2014). $V$-band observations by Wang \& Gao, continued during 10
days, show no essential light decay, which strongly suggests that
their observations cover the brightness maximum. However, this
suggestion may be incorrect in the case of plateau shape of the
light curve. Tartaglia et al. (2014a) carried out spectroscopic
observations and classified the object as a supernova impostor
similar to the prototypical SN~1997bs. SN~1997bs was classified by
Van Dyk et al. (2000) as a superoutburst of a very massive
luminous blue variable star, analogous to $\eta$~Car. In the
spectrum of the UGC~8246 transient taken on January~8, 2014,
Tartaglia et al. (2014a) found H$_\alpha$ emission with an
unresolved narrow component superimposed on broader wings (FWHM of
about 1800~km~s$^{-1}$). Taking into account scientific interest
to SN impostors, we continued spectroscopic and photometric
observations of this object.

\section{OBSERVATIONS AND DATA REDUCTION}

We obtained medium-resolution optical spectra of the transient PSN
J13100734+3410514 in UGC~8246 with the SCORPIO focal reducer
(Afanasiev \& Moiseev 2005) mounted on the 6-m BTA telescope of
the Special Astrophysical Observatory (SAO) on 2014 February~8 and
June~7. Seeing was measured as ${\rm FWHM} = 2\farcs5$ on February
8 and $2\farcs3$ on June~7. Additional photometry was performed
using the SAO 1-m Zeiss telescope and a CCD photometer with an
EEV42-40 chip on 2014 April 3, November 14 and 22. November
observations were carried out with the $V$ and $R_C$ filters at
good weather conditions with $\sim1^{\prime\prime}$ seeing, and
the total exposure times were 1200~s or longer.

In the photometric mode with the SCORPIO, we obtained CCD frames
with standard $B$, $V$ and Cousins $R$ filters. The color image
generated using these frames is shown in Fig.~1. It demonstrates
that the transient occurred in a spiral arm rich in star-forming
regions.

To perform relative photometry, we used the comparison star marked
``std'' in Fig.~1, its SDSS magnitudes being $ugriz =21.03,\
18.45,\ 17.34,\ 16.91,\ 16.68$. We transferred SDSS magnitudes to
$BVR_CI_C$ magnitudes by interpolation using $ugriz$ and $UBV$
AB$_{95}$ magnitudes of $\alpha$ Lyr as described in our previous
paper (Barsukova et al. 2012). As a result, the $BVR_CI_C$
magnitudes of the standard star were derived as (18.95, 17.97,
17.25, 16.49). All currently available photometry is presented in
Table~1. The light curves plotted using our data and all published
observations in the $V$ and $R_C$ filters are shown in Fig.~2.

\PZfig{7cm}{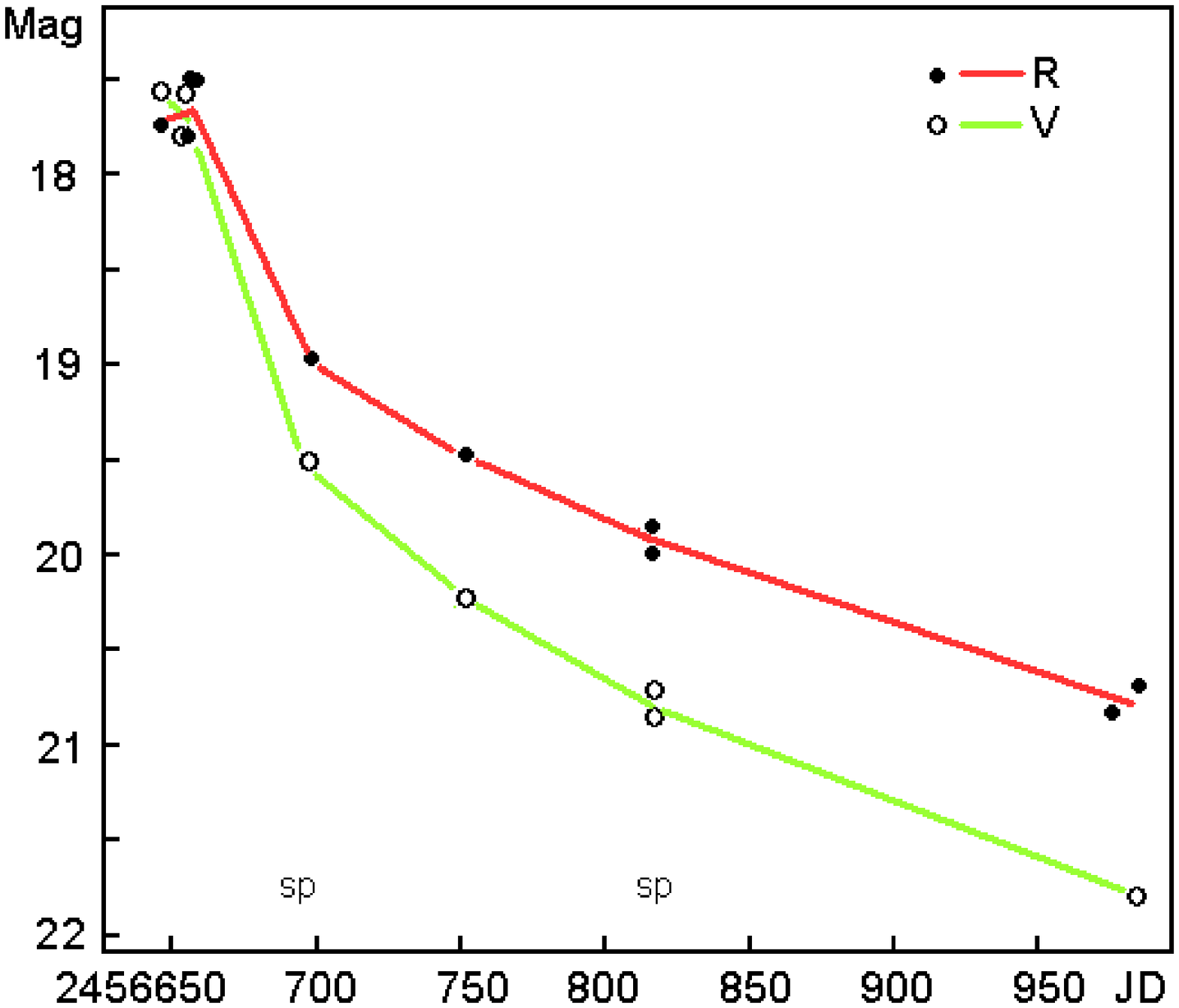}{ $V$ and $R_C$ light curves of the
transient PSN~J13100734+3410514 in UGC~8246. Dates of BTA/SCORPIO
spectroscopy are marked ``sp''. }

On February~8, 2014, the spectroscopic camera functioned in the
long-slit mode. First it was equipped with the VPHG 1200G grism
(nominal spectral range $\lambda\ 4000 - 5700$~\AA, resolution 5
\AA, dispersion 0.88 \AA~pixel$^{-1}$), and later it was replaced
with the VPHG 550G grism (spectral range $\lambda\ 3500 -
7200$~\AA, resolution 10~\AA, dispersion 2.1 \AA~pixel$^{-1}$).
Actual resolution measured using emission lines of night sky was
5.4~\AA\ for the VPHG 1200G grism and 14.6~\AA\ for VPHG 550G. In
the first spectrum, we found an essential contribution of a nearby
galactic H~II region in the brightest lines of the stellar
spectrum, so we took spectra with different position angles of the
slit. Slit locations at different telescope pointings are shown as
straight lines in Fig.~1 and they are listed in Table~2 for
individual spectra. On June~7, 2014, we used only the VPHG 550G
grism. Spectra were reduced using standard ESO MIDAS procedures
for the long-slit mode. Basic parameters of our spectra are
collected in Table~2. The blue and green spectral regions are
displayed in Fig.~3, and the whole optical spectra are presented
in Fig.~4. The wavelengths in these figures have been corrected
with respect to bright emission components of Balmer lines.

\PZfig{11cm}{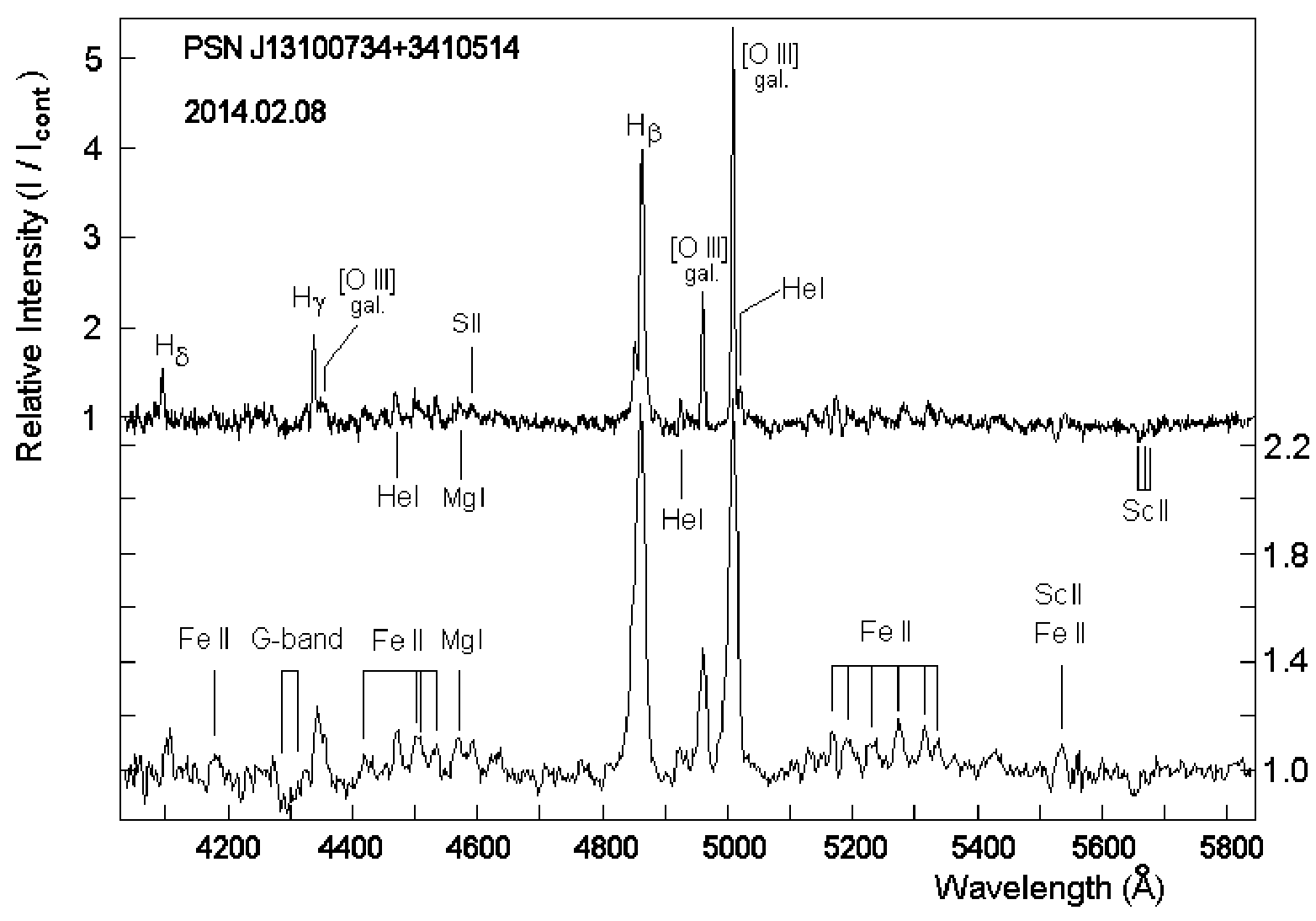}{ BTA/SCORPIO spectra of the UGC~8246
transient in the blue and green regions on February 8, 2014. Both
spectra are corrected for the redshift and presented for zero
velocity. Top: the higher resolution spectrum; bottom: the same
fragment of the lower resolution spectrum. }

\PZfig{10.5cm}{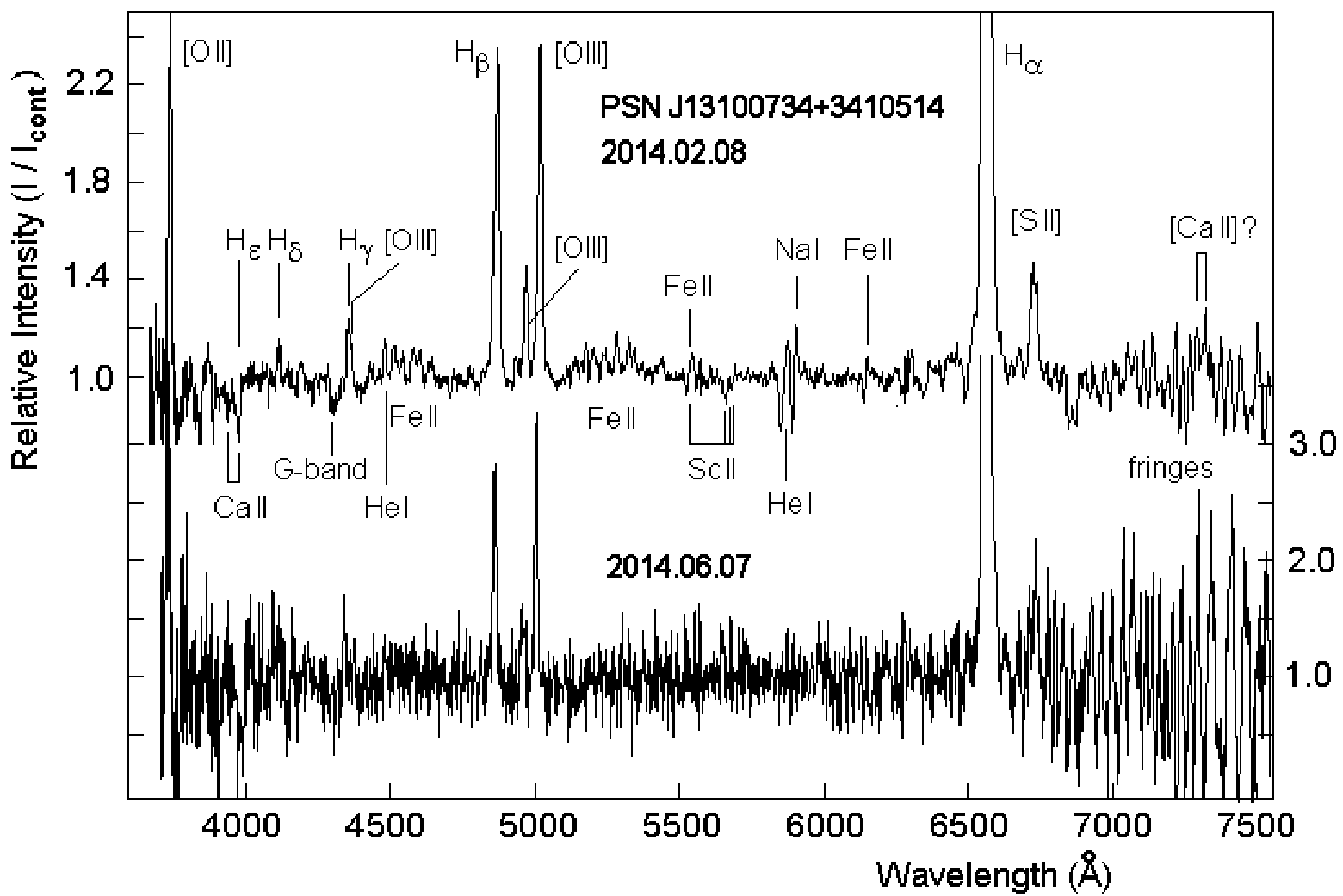}{A comparison of lower resolution
spectra of the UGC~8246 transient taken with BTA/SCORPIO on
February~8 and June~7, 2014. Both spectra are corrected for the
redshift. }

\PZfig{9.5cm}{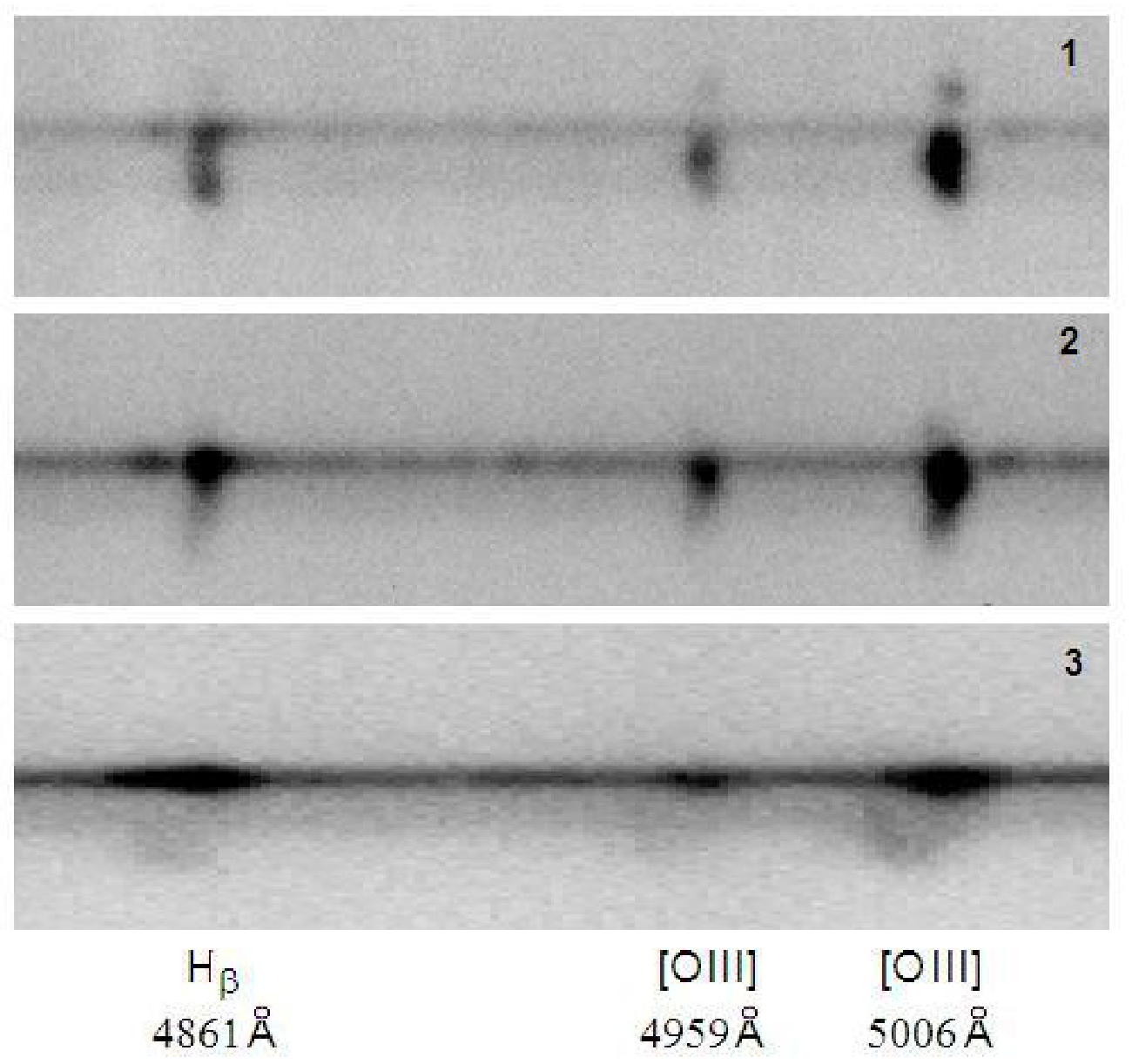}{ Two-dimensional BTA/SCORPIO
spectrograms of the UGC~8246 transient in the $\lambda\ 4800 -
5100$~\AA\ wavelength range for the three slit locations. The
numbers correspond to slit locations displayed in Fig.~1. [O~III]
emissions evidently belong to a close star-forming H~II region,
and the transient is located at its edge. At the same time, the
photos show that the radiation of the blue component belongs just
to the star, but the main component is contaminated by radiation
of the star-forming region which has the same radial velocity as
the star.}

\vspace{-1cm}

\section{RESULTS}

Analyzing published early non-homogeneous observations presented
in Table 1, one can suggest that the star had a $V-R_C$ color
index about zero at the outburst maximum. We do not know the
interstellar reddening in the galaxy UGC~8246 and therefore can
determine only the upper limit to the bolometric magnitude
($M_{bol} \le -13\fmm0$) and the lower limit to the effective
temperature ($\log T_e \ge 4.48$). Our February photometry taken
50~days after the maximum gave the following magnitudes and color
indices: $V = 19\fmm51$, $B-V = 0\fmm71$, $V-R_C = 0\fmm54$.
Reddening continued in June; on the 169th day after the light
maximum, $V-R_C$ was 0\fmm86. In November, on the 336th day at the
level of $V = 21\fmm8$, the star became even redder, with $V-R_C =
1\fmm09$. The star had a primary decay of $2\fmm0\ V$ in 50~days
and a nonlinear secondary decay both in the $V$ and $R_C$ bands.
By the 169th day after maximum, the speed of the decay decreased
to about 0\fmm7~$V$ in 100~days.

The spectrum of the UGC~8246 transient taken in February was rich
in emission lines. It resembled spectra of Fe~II-class novae in
the fireball stage (e.g., see Williams 2012; we used the finding
list of optical emission lines in his Table~2 for identification).
The lines we have identified are listed in Table~3. The strongest
emission lines in our spectra are Balmer lines, [O~II], and
[O~III]. Oxygen lines are evidently radiated in the nearby H~II
regions in the spiral arm of the galaxy. As seen in Fig.~5, the
regions of [O~III] emission are superimposed on the stellar
spectrum only partly. Otherwise, this figure shows that the main
component of H$_\beta$ emission mostly has a stellar origin. But
oxygen lines are located at the same radial velocity,
+920~km~s$^{-1}$, as the  strongest H$_\beta$ component. Direct
images taken in the $R_C$ filter in November confirm the
association of the transient with a compact H~II region, the
brightest part of which is located in 1\farcs0 east and 0\farcs5
south of the transient. These spectra and the direct images
definitely provide evidence that the explosion of this transient
has happened in the active star forming region rich in young
massive stars.

The equivalent widths of Balmer lines on February~8 were the
following: ${\rm EW}({\rm H}_\alpha) = -180$~\AA; ${\rm EW}({\rm
H}_\beta) = -37$~\AA; ${\rm EW}({\rm H}_\gamma) = -11$~\AA; ${\rm
EW}({\rm H}_\delta) = -3$~\AA. H$_\varepsilon$, in a blend with
Ca~II~H, form a deep absorption line. On June~7, the equivalent
widths changed to --350~\AA\ (H$_\alpha$), --41~\AA\ (H$_\beta$),
and --7~\AA\ (H$_\gamma$).

With the best resolution of 5.4~\AA, the H$_\beta$ line profile
looks complex (Fig.~6). The strongest narrow component has ${\rm
FWHM} = 320 \pm30$~km~s$^{-1}$ after correction for the spectral
resolution. The profile extends over ${\rm FWZI} =
2200$~km~s$^{-1}$. Additionally, there is another weak blue narrow
emission component in the profile, displaced by $-660$~km~s$^{-1}$
with respect to the main component and having approximately the
same half width, 320~km~s$^{-1}$. A dip between these components
is at the velocity of --380~km s$^{-1}$ and looks like an
absorption. Such profiles with an absorption component are rare in
spectra of SN impostors but do exist. With wings, an absorption
dip and a narrow component, the H$_\beta$ line profile of UGC~8246
transient resembles Balmer line profiles of the type IIP supernova
1994W in NGC~4041 (Sollerman et al. 1998), but that was a true and
extremely luminous supernova. A similar shape of Balmer line
profiles was observed by Humphreys et al. (2012) in the spectrum
of the peculiar type IIn SN or impostor 2011ht in UGC~5460. That
star had reached M$_V = -17\mm$ in maximum. However, the wing span
in the profiles of these two supernovae was 4~times larger than
that of the UGC~8246 transient. Also, wings of the H$_\beta$ line
profile are twice narrower than wings of the same profile of the
``calcium transient'' in NGC 5775 observed by us (Barsukova et al.
2012). There was no absorption component in the profile of that
transient. The narrow component in H$_\alpha$ profile of the UGC
8246 transient was poorly resolved with the VPHG 550G grism. The
line displayed a blue shoulder that confirmed the presence of a
similar blue narrow component at the velocity about
--660~km~s$^{-1}$. Explaining profiles of similar shape in Balmer
lines of SN~1994W, Chugai et al. (2004) identified their
components with a photoionized expanding circumstellar envelope,
shocked cool gas in the forward post-shocked region, and multiple
Thompson scattering in the envelope. Probably cool gas is
responsible for the absorption component in the profiles of Balmer
lines. The models gave a rather deep absorption component in a
case of homologous expansion of the ejecta without post-explosion
radiative acceleration but did not take into consideration light
echo in the circumstellar medium.

\PZfig{7.0cm}{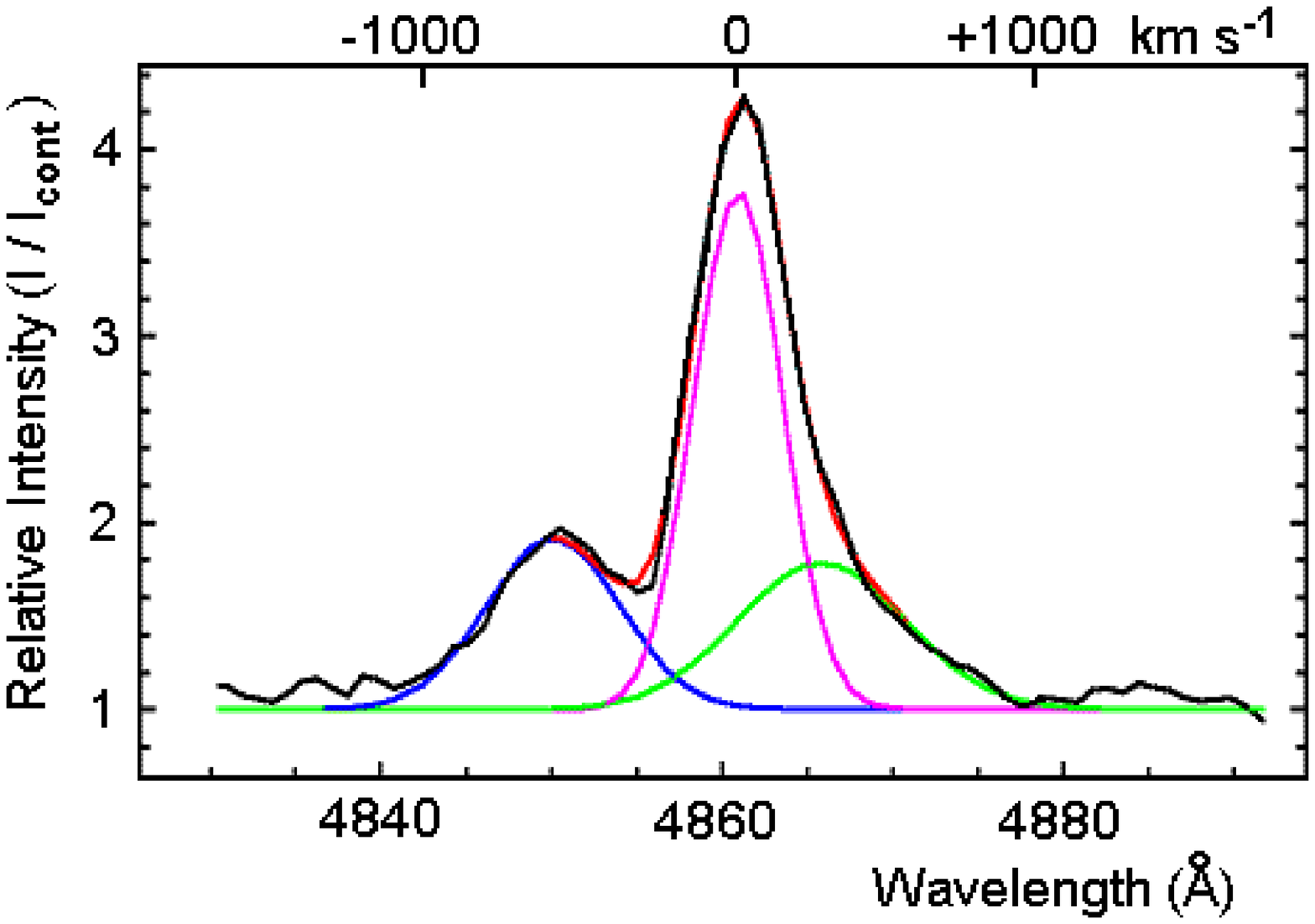}{ The H$_\beta$ line profile in the
spectrum with the resolution of 5.4 \AA\ taken on 2014 February~8
(black curve). Gaussian components are shown in blue, pink, and
green colors. The model profile defined as the sum of the three
components is plotted as the red curve. }

One can expect that the light echo of an LBV explosion expanding
through the circumstellar envelope formed by stellar wind of the
progenitor should influence the Balmer line profiles. An external
observer can see only those illuminated dust particles or emissive
excited atoms that are located on the surface of the ellipsoid
with the flare source in one focus, and the observer in the other
one. In this case, the total light path from the source to a
particle and of the reflected light from the particle to the
observer is the same for all visible particles at the time $t$
after the flare, namely $c\cdot t$. During the first weeks after
the outburst, the ellipsoid is narrow and elongated along the line
of sight. Then the photoionization front in the expanding gaseous
envelope visible for the observer in the first weeks includes only
the part of the envelope that is located along the line of sight
and moving to the observer. The light from the excited gas at the
opposite side of the envelope and from other parts of it has not
yet reached the observer. Thus, in the first weeks, we expect to
have more negative velocity of the main emission component
relative to its velocity during the last stages. To test the
assumption on the light echo, we measured velocities of peaks in
the H$_\alpha$ and H$_\beta$ profiles in the spectrum by Tartaglia
et al. (2014a) taken on January 8, 2014 and found 820 and
850~km~s$^{-1}$, respectively, against 970 and 920~km~s$^{-1}$ in
our February~8 spectra. So the velocity difference is really
present, and is equal to $-110 \pm40$~km~s$^{-1}$. Later, the
visible front of ionization will get to the opposite side of the
wind envelope, with velocities directed away from the observer,
and the central-peak velocity will increase. Our spectrum taken on
June~7 gives larger velocities, 1060~km~s$^{-1}$ for H$_\alpha$
and 960~km~s$^{-1}$ for H$_\beta$ (these values were corrected
with respect to galactic [O~III] lines). Note that the presence of
a negative excess in the velocity of the main emission component
of $H_\alpha$ on January~8, 2014 confirms the hypothesis that the
transient was discovered at an early stage of the outburst and
testifies against a plateau stage. Thus, the effects of light echo
in Balmer-line profiles in the spectra of SN~impostors are strong
and should be taken into account in the spectral line modeling.

We have tried to separate components of the H$_\beta$ line profile
(Fig.~6) using the MIDAS procedure \hbox{DEBLEND/LINE.} These
components were fitted with Gaussian functions. The fit with a
broad component along with two narrow ones represents the total
profile badly, especially to the red side. The observed curve runs
evidently above the model one. An attempt to fit the profile
including one absorption component represents the data better but
results in a very broad main component and a broad absorption
which eats up half of main emission width. Such a model is hard to
explain physically. But the fit without any broad component and
with three narrow ones is the best. In short, the red excess in
the profile can be successfully fitted with an additional
Gaussian. Parameters of the components in this model are the
following: for the blue one, $v_R = -670 \pm 15$~km~s$^{-1}$,
${\rm FWHM} = 570 \pm40$~km~s$^{-1}$, ${\rm EW} = -8.9$~\AA; for
the central one, $v_R = 0 \pm20$~km~s$^{-1}$ (adopted), ${\rm
FWHM} = 376\pm15$~km~s$^{-1}$, ${\rm EW} = -18.0$~\AA; for the red
one, $v_R = +300\pm 70$~km~s$^{-1}$, ${\rm FWHM} = 715
\pm85$~km~s$^{-1}$, ${\rm EW} = -9.7$~\AA. Note that the FWHM
values in this model include the instrumental profile. The fitting
and its three Gaussian components are shown in Fig.~6.

We have an alternative explanation of the H$_\beta$ line profile
based on this fitting. A separate narrow blue component may be the
radiation of a frontal lobe of bipolar ejecta thrown out in the
direction of the observer, while the opposite receding lobe is
partly covered by the volume of the approaching lobe. The
non-covered radiation of the receding lobe is visible in the red
part of the profile. Note that a contribution from the H~II region
may be partly added to the red component (see Fig.~5). Such a
bipolar eruption formed the ``Homunculus Nebula'' of $\eta$~Car in
the mid-1800s explosion. The lobes in the Homunculus are known to
be expanding radially with velocities in the range of
650--700~km~s$^{-1}$ (Davidson \& Humphreys 1997) which is in
agreement with 670~km~s$^{-1}$ for the UGC~8246 transient.
However, the UGC~8246 transient outburst was not so long in
duration, only 200~days vs. 20~years for $\eta$ Car.  In this
case, the dip between two narrow components in the profile of
H$_\beta$ cannot be treated as an absorption.

The rich weak-line spectrum indicates a dense environment
containing evolved material ejected from the progenitor by the
stellar wind in the super-Eddington regime. There are many Fe~II
emission lines in the spectrum, the strongest ones have P~Cyg
profiles. The absorption component of the strongest Fe~II
5169~\AA\ line is located at the velocity of $-400$~km~s$^{-1}$.
Forbidden emissions are absent in the stellar spectrum, but they
are present in the spectrum of the galaxy environment. The stellar
spectrum indicates low excitation and absence of a hot ionizing
source. He~I 5876~\AA\ is very strong, it has a P~Cyg profile; its
absorption component is centered at the velocity of
--1500~km~s$^{-1}$ and expands to --1850~km~s$^{-1}$ with respect
to the center of the emission component. The center of the
emission component is displaced by --400~km~s$^{-1}$ against main
components of the Balmer lines. The profile looks strange and
unlike any other lines, and there is an assumption that it is
being formed entirely in the ejecta, not in the wind. He~I
4471~\AA\ is also strong, so we have to identify emission lines at
4922 and 5016~\AA\ as He~I lines, though they can contain some
contribution from close Fe~II components. The Na~I D$_2$D$_1$
blend has a P~Cyg profile and is as strong as He~I 5876~\AA. The
absorption component of Na~I is displaced to the blue side from
the emission component's center by 700~km~s$^{-1}$.

The Ca~II H \& K lines are in absorption. In the spectra, we also
identify the Mg~I 4571~\AA\ intercombination line that may be
formed in massive and rarefied nebulae, and sometimes occurs in
ejecta of red novae (Goranskij \& Barsukova 2007). Recently
Tartaglia et al. (2014b) found absorption lines of Ba~II and Sc~II
in the spectra of the SN impostor 2007sv that exploded in the
galaxy UGC 5929; these species are not met in the spectra of
classical novae. The presence of Sc~II, Mg~II, and Si~II was noted
in the spectra of SN~1994W (Chugai et al. 2004). We tried to find
these lines in our spectra of the UGC 8246 transient, and the
attempt seems to be successful only for Sc~II. There is an obvious
depression at $\lambda$5640--5684~\AA\ that cannot be explained by
absorption of other chemical elements except Sc~II. The line of
Sc~II 5526 \AA\ can contribute to the absorption component of the
Fe~II 5535~\AA\ P~Cyg-type profile. We also identified a strong
absorption feature located towards short wavelengths from
H$_\gamma$ at $\lambda$4294--4315~\AA, like the G~band in early
G-type stars. This molecular CH~band was found in the spectrum of
the $\eta$~Car-type transient SN~2011ht by Humphreys et al.
(2012).

\PZfig{7.0cm}{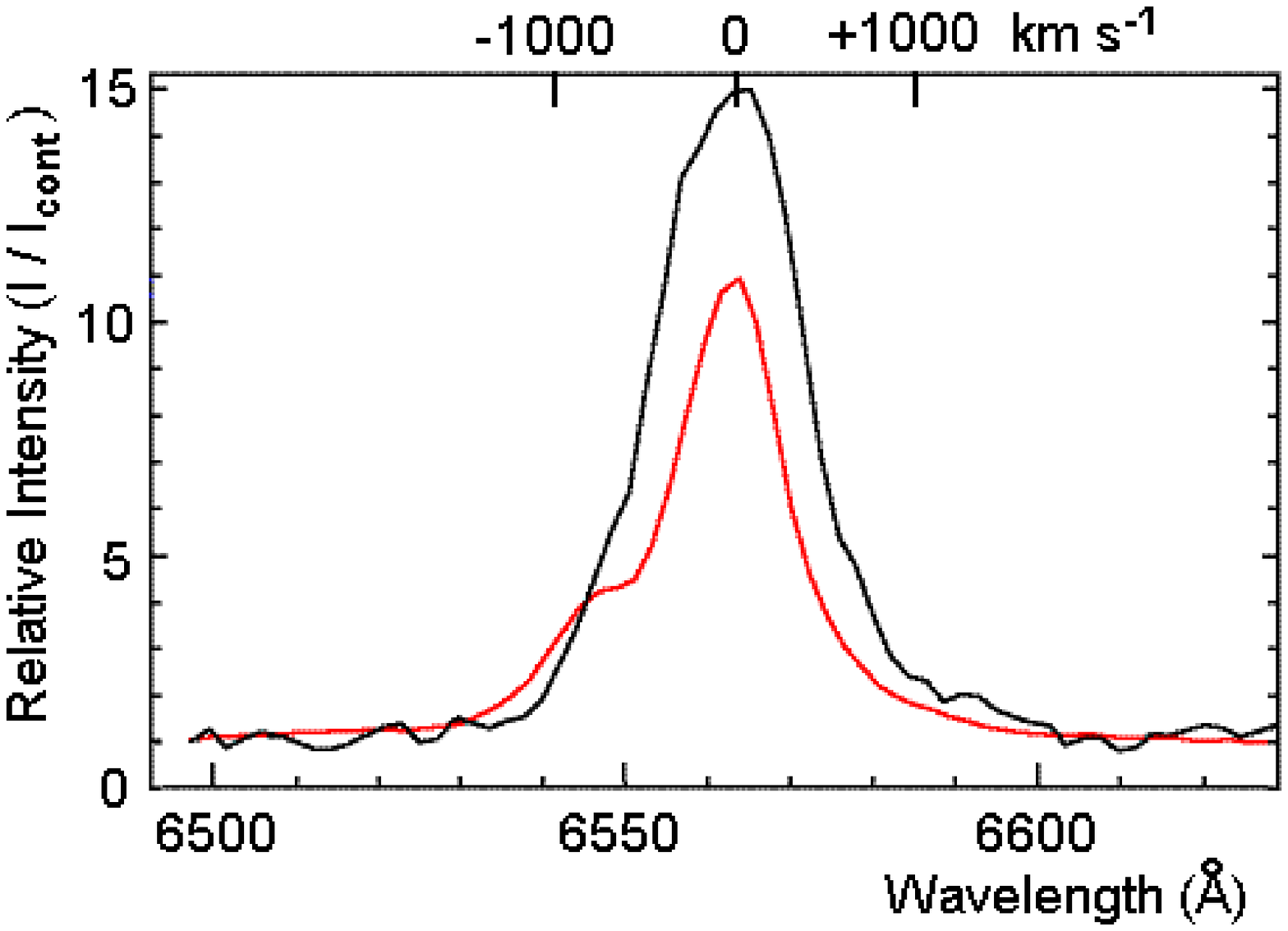}{ Evolution of the H$_\alpha$ line
profile for 119 days between 2014 February 8 (red curve) and 2014
June 7 (black curve) in the BTA/SCORPIO normalized spectra with
the resolution of 14.6~\AA. }

In the low-resolution noisy spectrum taken on June~7 (on the 169th
day after maximum), when the object's brightness was at the level
of 20\fmm8 in the $V$ band, only the strongest lines of the
transient and the galaxy are visible. The Balmer lines definitely
belong to the transient: the star's continuum can be distinguished
in the spectrum, and the stellar image is visible in the direct
frames. Figure~7 compares the H$_\alpha$ line profiles in the
February~8 and June~7 spectra. The narrow H$_\alpha$ line became
noticeably broader on June~7 than on February~8. In spite of low
resolution, the red and blue components stay visible in the
profile as shoulders. Note that these components may be the
radiation of bipolar ejecta. The half width of the profile
increased from 300 to 700~km~s$^{-1}$. These estimates were
corrected for the spectral resolution, and their accuracy is about
50~km~s$^{-1}$. Both profiles are plotted in intensity units
relative to the stellar continuum. The star faded between these
dates by 1\fmm0 in the $R_C$ band. The contribution of continuum
in the $R_C$ band changed between these dates from 82 to 78 per
cent. As the H$_\alpha$ flux declined slower than the continuum,
the equivalent width of the H$_\alpha$ line increased its absolute
value from 180 to 350~\AA. We explain the broadening of the
H$_\alpha$ line by two reasons. First, it is the radiative
acceleration of the stellar wind. The second reason is the
expansion of visible volume of the photoionized region of stellar
wind due to expanding light echo.

The observations of LBV-type transient remnants with the {\it
Hubble} space telescope or {\it Spitzer} infrared space telescope
show that the LBV stars survive in their explosions. Some of them
undergo repeated explosions. Thus, future observations of the
UGC~8246 transient are encouraged.

\section{CONCLUSIONS}

We present the results of our spectroscopy and CCD photometry of
the intermediate-luminosity red transient PSN~J13100734+3410514 in
UGC~8246 performed in February and June, 2014 with the Russian 6-m
BTA telescope and SCORPIO spectral camera and follow-up CCD
photometry with the SAO 1-m telescope.

The transient showed a  number of characteristics typical of LBV
stars with giant eruptions. Namely, the amplitude of the outburst
exceeded three magnitudes; the bolometric magnitude at maximum was
evidently above the Humphreys--Davidson limit (the upper boundary
of the supergiant luminosities in the $M_{bol} - T_{eff}$ diagram;
see Humphreys \& Davidson 1979); a significant reddening of the
spectrum occurred during the outburst decline; the spectrum is
typical of LBV stars; moderate wind velocities are characteristic
of LBV stars; a possible bipolar ejecta signature was seen in the
Balmer line profiles. There is an evidence for a light echo of the
outburst expanding in the surrounding gaseous medium after the
outburst. We assume, like other researchers, that this transient
was an explosion of a high-mass LBV star.

\textit{Acknowledgments.}

This work was supported by the Russian Foundation for Basic
Research (RFBR) with the grants 13-02-00885 and 14-02-00759.
E.A.B. thanks the President of the Russian Federation for the
grant supporting the Leading scientific school NSh-4308.2012.2.
A.F.V. thanks the President for grants for young PhD researchers
MK-6686.2013.2 and MK-1699.2014.2 and the RFBR for the grant
12-02-31548. We thank Drs. D.Yu~ Tsvetkov and O.V.~Maryeva for
discussion. Observations with the Russian 6-m telescope are
financially supported by the Ministry of Education and Science of
Russian Federation (agreement No. 14.619.21.004, project
identifier RFMEFI61914X0004). This research has made use of the
Sloan Digital Sky Survey and of the NASA/IPAC Extragalactic
Database (NED).

\references

Afanasiev, V.L., Moiseev, A.V., 2005, {\it Astronomy Letters},
{\bf 31}, 194

Brimacombe, J., 2014,\\
http://spacefinalfrontier.blogspot.ru/2014/02/luminous-blue-variable-psn.html

Barsukova,~E.A., Fabrika,~S.N., Goranskij,~V.P., Valeev,~A.F.,
2012, {\it Variable Stars}, {\bf 32}, No.~2

Barsukova,~E.A., Goranskij,~V.P., Valeev,~A.F., Zharova,~A.V.,
2014, {\it Astrophys. Bull.}, {\bf 69}, 67

Berger, E., Soderberg,~A.M., Chevalier,~R.A. et al., 2009, {\it
Astrophys. J.}, {\bf 699}, 1850

Chugai,~N.N., Blinnikov,~S.I., Cumming,~R.J., et al., 2004, {\it
MNRAS}, {\bf 352}, 1213

Corcoran,~M.F., Ishibashi, K., 2012, in {\it Eta Carinae and the
Supernova Impostors}, R.~Davidson \& R.M.~Humphreys (eds.),
Springer, New York, p.195

Davidson, K., Humphreys, R.M., 1997, {\it Ann. Rev. Astron. \&
Astrophys.}, {\bf 35}, 1

Goranskij,~V.P., Barsukova,~E.A., 2007, {\it Astron. Reports},
{\bf 51}, 126

Humphreys,~R.M., Davidson,~K., 1979, {\it Astrophys. J.}, {\bf
232}, 409

Humphreys,~R.M., Davidson,~K., Jones,~T.J., et al., 2012, {\it
Astrophys. J.}, {\bf 760}, 93

Ishibashi,~K., Corcoran,~M.F., Davidson,~K., et al., 1999, {\it
Astrophys. J.}, {\bf 524}, 983

Kashi, A., Soker, N., 2010, {\it arXiv}:1011.1222

Kashi, A., 2010, {\it AIP Conference Proc.}, {\bf 1314}, 55

Margutti,~R., Milisavljevic,~D., Soderberg,~A.M. et al., 2014,
{\it Astrophys. J.}, {\bf 780}, 21

Martini,~P., Wagner,~R.M., Tomaney,~A. et al., 1999, {\it Astron.
J.}, {\bf 118}, 1034

Mauerhan,~J.C., Smith,~N., Filippenko,~A.V. et al., 2013, {\it
MNRAS}, {\bf 430}, 1801

Mehner,~A., Davidson,~K., Ferland,~G.J. et al., 2010, {\it
Astrophys. J.}, {\bf 710}, 729

Owocki,~S.P., Shaviv,~N.J. 2012, in {\it Eta Carinae and the
Supernova Impostors}, K.~Davidson \& R.M.~Humphreys (eds.),
Springer, New York, p.275

Smith,~N., Ganeshalingam,~M., Chornock,~R. et al., 2009, {\it
Astrophys. J.}, {\bf 697}, L49

Sollerman,~J., Cumming,~R.J., Lundquist,~P., 1998, {\it Astrophys.
J.}, {\bf 493}, 933

Tartaglia,~L., Pastorello,~A., Benetti,~S. et al., 2014a, {\it
Astronomer's Telegram}, No.~5737

Tartaglia,~L., Pastorello,~A., Taubenberger,~S. et al., 2014b,
{\it arXiv}:1406.2120

Tylenda,~R., Soker,~N. 2006, {\it Astron. \& Astrophys.}, {\bf
451}, 223

Tylenda,~R., Hajduk,~M., Kaminski,~T. et al., 2011, {\it Astron.
\& Astrophys.}, {\bf 528}, 114

Van Dyk,~S.D., Schuyler,~D., Peng,~C.Y. et al, 2000, {\it Publ.
Astron. Soc. Pacific}, {\bf 112}, 1532

Wang,~B., Gao,~X., 2013, CBAT\\
(http://www.cbat.eps.harvard.edu/unconf/followups/J13100734+3410514.html)

Williams,~R., 2012, {\it Astron. J.}, {\bf 144}, 98

\endreferences

\begin{table}
  \caption{Available photometry of PSN J13100734+3410514}
  \smallskip
  \begin{center}
\begin{tabular}{lllcl}
\hline
\\
Date       & JD hel.      & mag & Filter & Source  \\
           & 2400000+     &     &        &         \\
\\
\hline\\
2013.12.20  & 56647       & 17.75 & $R$     &  Bin Wang, Xing Gao  (CBAT)\\
2013.12.20  & 56647       & 17.6  & $V$     &  Bin Wang, Xing Gao  (CBAT)\\
2013.12.28  & 56655       & 17.8  & $R$     &  L. Elenin (CBAT)          \\
2013.12.28  & 56655       & 17.8  & $V$     &  Bin Wang, Xing Gao  (CBAT)\\
2013.12.29  & 56656       & 17.6  & $V$     &  Bin Wang, Xing Gao  (CBAT)\\
2013.12.30  & 56657       & 17.5  & $R$     &  F. Luppi, L. Bussi (CBAT)\\
2013.12.31  & 56658       & 17.51 & $V$     &  Brimacombe (2013) $\pm0.12$\\
2014.02.08  & 56697.5559  & 19.51 & $V$     &  BTA/SCORPIO \\
2014.02.08  & 56697.5574  & 20.22 & $B$     &  BTA/SCORPIO \\
2014.02.08  & 56697.5589  & 18.97 & $R_C$ &  BTA/SCORPIO \\
2014.04.03  & 56751.3947  & 20.23 & $V$     &  Zeiss-1000/EEV42-40\\
2014.04.03  & 56751.4096  & 19.67 & $R_C$ &  Zeiss-1000/EEV42-40\\
2014.04.03  & 56751.4138  & 20.51 & $B$     &  Zeiss-1000/EEV42-40\\
2014.06.07  & 56816.2918  & 19.85 & $R_C$ &  BTA/SCORPIO \\
2014.06.07  & 56816.2945  & 20.71 & $V$     &  BTA/SCORPIO \\
2014.06.07  & 56816.2965  & 20.85 & $V$     &  BTA/SCORPIO \\
2014.06.07  & 56816.2984  & 19.99 & $R_C$ &  BTA/SCORPIO \\
2014.11.14  & 56975.6144  & 20.81 & $R_C$ &  Zeiss-1000/EEV42-40\\
2014.11.22  & 56983.5790  & 21.83 & $V$     &  Zeiss-1000/EEV42-40\\
2014.11.22  & 56983.5963  & 20.67 & $R_C$ &  Zeiss-1000/EEV42-40\\
\\
\hline
\end{tabular}
\end{center}
\end{table}

\begin{table}
  \caption{Spectra of PSN J13100734+3410514 taken with BTA/SCORPIO$^*$)}
  \smallskip
  \begin{center}
\begin{tabular}{llcllll}
\hline
\\
Time of mid-  & Exposure, &$\lambda$ & S/N,& Grism  & Resolution, & Slit   \\
exposure, UT   &    \phantom{22}s  &  \AA     &    &        &    \phantom{22}\AA    & location\\
\\
\hline\\
2014 Feb 08.984  &  1200  &  4053--5848       & \phantom{2}7   & VPHG1200G &  \phantom{2}5.4 & 1\\
2014 Feb 09.027  &  2400  &  4053--5848       & 15  & VPHG1200G &  \phantom{2}5.4 & 2\\
2014 Feb 09.106  &  2400  &  3644--7906       & 25  & VPHG550G  & 14.6 & 3\\
2014 Jun 06.970  & \phantom{2}600  &  3716--7905       & \phantom{2}2   & VPHG550G  & 14.6 & 3\\
\\
\hline \multicolumn{7}{l}{\footnotesize $^*)$ Original normalized
spectra in ASCII code are available in the Internet at
http://jet.sao.ru/$\sim$bars/spectra/psn1310/ }
\end{tabular}
\end{center}
\end{table}

\begin{table}
  \caption{Line identifications}
  \smallskip
  \begin{center}
\begin{tabular}{lcl}
\hline
\\
Element       &$\lambda_0$, \AA & Notes \\
\\
\hline\\
\lbrack O II\rbrack           & 3727       & blend, galaxy \\
Ca II                         & 3934       & absorption  \\
Ca II                         & 3968       & blend with H$_\epsilon$  \\
H$_\epsilon$                  & 3970       & absorption  \\
H$_\delta$                    & 4102       & P Cyg       \\
Fe II                         & 4179       &             \\
G band                        & 4294--4315 & CH molecular absorption  \\
H$_\gamma$                    & 4340       & P Cyg       \\
\lbrack O III\rbrack          & 4363       & galaxy      \\
Fe II                         & 4417       &             \\
He I                          & 4471       &             \\
Fe II                         & 4508       &             \\
Fe II                         & 4515       &             \\
Fe II                         & 4534       &             \\
Mg I\rbrack                   & 4571       & intercombination line \\
S II?                         & 4591       & blend       \\
H$_\beta$                     & 4861       & P Cyg       \\
He I                          & 4922       & P Cyg       \\
\lbrack O III\rbrack          & 4959       & galaxy      \\
\lbrack O III\rbrack          & 5007       & galaxy      \\
He I                          & 5016       &             \\
Fe II                         & 5169       & P Cyg       \\
Fe II                         & 5198       &             \\
Fe II                         & 5235       &             \\
Fe II                         & 5276,5284  & blend       \\
Fe II                         & 5317       & P Cyg       \\
Fe II                         & 5333,5337  & blend       \\
Fe II                         & 5535       &             \\
Sc II                         & 5526       &             \\
Sc II                         & 5667       & blend, absorption  \\
He I                          & 5876       & P Cyg       \\
Na I D$_2$,D$_1$              & 5889,5892  & P Cyg       \\
Fe II                         & 6148       & P Cyg       \\
H$_{\alpha}$                  & 6563       & emission    \\
\lbrack S II\rbrack           & 6716,6731  & galaxy      \\
\lbrack Ca II\rbrack?         & 7296,7329  & weak, blend?\\
\\
\hline
\end{tabular}
\end{center}
\end{table}

\end{document}